\documentclass[%
aps,
prl,
nofootinbib,
nobibnotes,
amsmath,
amssymb,
twocolumn
]{revtex4-2}
\usepackage{amssymb}
\usepackage{physics}
\usepackage{graphicx}
\usepackage{amsmath}
\usepackage{amsthm}
\usepackage{amsfonts}
\usepackage{bbm}
\usepackage{soul}
\graphicspath{{./imgs}}

\usepackage{xcolor}
\usepackage{hyperref}
\usepackage[caption=false, subrefformat=parens]{subfig}

\usepackage{bbold}
\usepackage{comment}

\newcommand{\expect}[1]{\left\langle {#1} \right\rangle}

\renewcommand\section[1]{\emph{#1}.---}

\begin{document}

\title{Theory of anomalous Hall effect from screened vortex charge in a phase disordered superconductor}

\author{Jay D. Sau}
\email[]{jaydsau@umd.edu}
\author{Shuyang Wang}
\affiliation{Condensed Matter Theory Center and Joint Quantum Institute, Department of Physics, University of Maryland, College Park, Maryland 20742-4111, USA}

\date{\today}

\begin{abstract}
Motivated by recent experiments showing evidence for chiral superconductivity in an anomalous Hall phase of tetralayer graphene, we study the relation between the normal state anomalous Hall conductivity  and that in the phase disordered state above the critical temperature of the superconductor. By a numerical calculation of superconductivity in an anomalous Hall metal, we find that a difference in vortex and antivortex charge is determined by the Fermi surface Berry phase. Combining this with the vortex dynamics in a back-ground supercurrent leads to a Hall response in the phase disordered state of the superconductor that is close to the normal state anomalous Hall response. However, using a gauge-invariant superconducting response framework, we find that while vortex charge is screened by interactions, the screening charge, after a time-delay,  reappears in the longitudinal current. Thus, the dc Hall conductivity in this phase, instead of matching the screened vortex charge, matches the ac Hall conductance in the superconducting and normal phase, which are similar.
\end{abstract}


\maketitle

\section{Introduction}%
Multilayer graphene has recently shown evidence of a number of novel phases 
that can be tuned by gate voltage, magnetic field, temperature and displacement field. These phases include several superconducting phases in twisted systems~\cite{Cao2018correlated,Cao2018unconventional,Yankowitz2019tuning,Lu2019superconductors,Hao2021electric,Park2021tunable,Cao2021pauli,Liu_2022isospin,zhang2021ascendance,park2021magic,Burg2022emergence} and also in non-twisted systems ~\cite{Zhou2021half,Zhou2021superconductivity,Zhou2022isospin} some of which are spin triplet. 
More interestingly, a recent experiment~\cite{han2024signatures} has 
provided evidence of chiral superconductivity in an anomalous Hall metal phase, which is quite close to systems that have shown 
quantum anomalous Hall as well as fractional quantum anomalous Hall phases~\cite{han2024correlated}. 
The occurrence of superconductivity in close proximity to correlated phases has led to many theoretical proposals 
for the mechanism of superconductivity, some based on strong correlation
~\cite{chou2024intravalley,geier2024chiral} and others based 
on the proximity to correlated topological states~\cite{kim2024topological,shi2024doping} .
\begin{figure}
    \centering
    \includegraphics[width=\columnwidth]{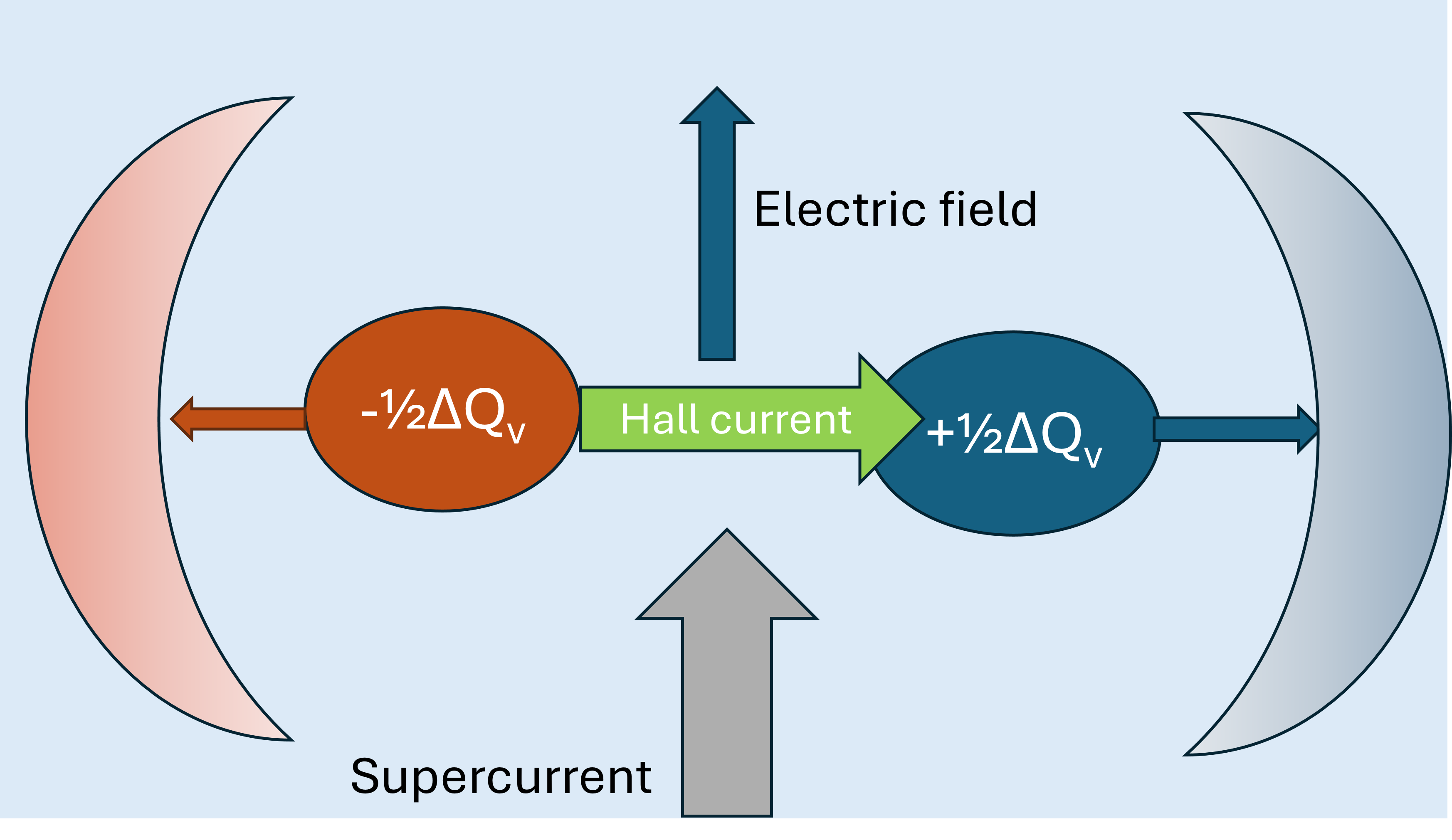}
    \caption{Schematic for the origin of anomalous Hall in a 2D superconductor above the BKT transition i.e. so-called phase disordered state. The local supercurrent applies a Magnus force~\cite{minnhagen1987two} on the vortex-antivortex pair (red and blue discs) in opposite directions. This leads to diffusive motion of the vortices along $\pm \hat{x}$~\cite{bardeen1965theory}. The motion of the vortices corresponds to an electric field along $y$, which is the origin of dissipative transport in the phase disordered state. 
    The difference in vortex-anti-vortex charge $\Delta Q_v = Q_- - Q_+$ that we will numerically show to be related to the normal state anomalous Hall coefficient lead to a Hall current (green arrow) orthogonal to the electric field. We find that many-body screening as well as the far-field phase winding of vortices (shown as the crescents of moving charge) conspire to match the measured dc Hall conductance from vortices to the bulk ac Hall conductance.}
    \label{fig:schematic}
\end{figure}

The peculiar characteristics of the superconducting state such as the pair density wave-character \cite{geier2024chiral}, chiral nature~\cite{kim2024topological} as  well as the Berry curvature of the band are likely to lead to interesting phenomelogical 
aspects that are quite independent from the origin of the superconductivity. In fact, quantum geometry, which is a generalization 
of Berry curvature, has already been shown to have a significant modification of the superfluid stiffness~\cite{Peotta2015superfluidity,Liang2017band,Hofmann2020superconductivity,Torma2022superconductivity,herzogarbeitman2022manybody,Huhtinen2022revisiting,Mao2023diamagnetic,Hofmann2023superconductivity}
even in the absence of Berry curvature.
Berry curvature of a band leads to more interesting behavior in the form of anomalous Hall conductance~\cite{karplus1954hall,jungwirth2002anomalous,haldane2004berry} in the normal state. This leads to the natural question 
about how such an anomalous Hall conductance would manifest in a chiral superconducting phase. 
In fact, chiral superconductivity by itself has been suggested to support an ac Hall response~\cite{yakovenko2007theory,mineev2007broken}. This chiral response has been conjectured to have a number of interesting 
consequences such as fractional charge and angular momentum of vortices~\cite{goryo2000vortex}. However, a gauge invariant treatment 
of screening effects by the background condensate leads to an effective Chern-Simons theory where the chiral conductivity from purely 
chiral superconductivity at low wave-vectors is suppressed~\cite{Lutchyn2008gauge,roy2008collective}.
The evidence for chiral superconductivity in a purely two dimensional anomalous Hall metal phase~\cite{kim2024topological}, where 
screening effects are reduced and phase disordered superconductivity is seen, motivates us to revisit the question of chiral response and vortex properties of such phases.

In this work, we will study the effect of anomalous Hall conductivity in the normal state on the phase disordered state in the superconductor above the Berezinski-Kosterlitz-Thouless (BKT) transition~\cite{minnhagen1987two}. As shown in Fig.~\ref{fig:schematic}, the difference in charge densities in the cores of vortices and anti-vortices can lead to an anomalous Hall contribution to the current. We argue that such a contribution can arise from a gauge invariant effective action of a superconductor~\cite{altland2010condensed} that includes a Hall response and numerically check that such a contribution indeed appears in a simple model. We then show using the gauge-invariant response that while the vortex charge is screened by many-body interactions, the  
dynamical screening cloud (crescents in Fig.~\ref{fig:schematic}) combine so that the vortex generated dc Hall conductivity is the same as the ac Hall response.

\section{Effective action of a Hall superconductor}\label{sec:Effectiveaction}%
To describe a superconductor, we introduce a fluctuating field $\phi(r,t)$, which will represent the phase of the symmetry breaking order parameter. The gauge transformation properties of $\phi$ are such that the shifted gauge potentials $b_\alpha=A_\alpha-\partial_\alpha \phi$ are gauge-invariant degrees of freedom, which is the essence of gauge fields acquiring mass~\cite{altland2010condensed}. As elaborated in Appendix. A
for the case of an electronic system similar to tetra-layer graphene assuming screened Coulomb interactions, the field $\phi$ can be microscopically defined as the phase of a Hubbard-Stratonovich field associated with the pairing interaction and appears as fields $b_\alpha$ in the effective action following the Hubbard-Stratonovich decomposition. 
Expanding this effective action to lowest order in the fields $b_\alpha$ in Fourier space $(q,\omega)$ components $e^{i (q\cdot r-\omega t)}$, leads to the expression:
\begin{align}\label{eq:Seff}
    &S_{eff}=\sum_{q,\omega} b_\alpha(q,\omega) b_\beta^*(q,\omega)K^{(\alpha,\beta)}(q,\omega)
\end{align}
where $K^{(\alpha,\beta)}(q,\omega)$  is a Hermitean matrix i.e. $K^{(\alpha,\beta)*}(q,\omega)=K^{(\beta,\alpha)}(q,\omega)$ for a gapped superconductor.
This together with the reality of $b_\alpha$ i.e  $b_\alpha(q,\omega)^*=b_\alpha(-q,-\omega)$ implies that $S_{eff}$ is non-dissipative. 
The ac current (and charge) in the superconductor can be obtained 
as functional derivatives of the action i.e. 
\begin{align}
    &j_\alpha(q,\omega)=\frac{\delta S_{eff}}{\delta A_\alpha^* (q,\omega)}=K^{(\alpha,\beta)}(q,\omega)b_\beta(q,\omega),\label{eq:linresp}
\end{align}
so that $K$ can be viewed as part of the electromagnetic response coefficients of the superconductor~\cite{schrieffer2018theory}.
The reality of the current further requires $K^{(\alpha,\beta)*}(q,\omega)=K^{(\alpha,\beta)}(-q,-\omega)$. 
Expanding $K^{(\alpha,\beta)}(q,\omega)$ to lowest non-zero order in $q,\omega$ consistent with these constraints (see Appendix. B for detailed form),
substituting $K$ into Eq.~\ref{eq:Seff} and Fourier transforming to space and time, the effective action $S_{eff}$ can be written as a gradient expansion:
\begin{align}\label{eq:SE}
    & S_{eff}=\int [-C_1 b_0^2+C_2 b^2-\{C_3 b-C_4 (\hat{z}\times b) \}\cdot \nabla b_0 \nonumber\\
    &+C_5 (\hat{z}\times b)\cdot \dot{b}].
\end{align}
The first two coefficients $C_1$ and $C_2$ are the superfluid compressibility and  stiffness respectively.  In the case $C_4=C_5$, the last two terms Fourier transform to $b_\alpha^*(q,\omega)\epsilon_{z\alpha\beta}(i q_\beta b_0-i\omega b_\beta)=i\epsilon_{\alpha\beta\gamma}b_\alpha^*(q,\omega)q_\beta b_\gamma(q,\omega)$ is exactly the Chern-Simons term in the superconductor~\cite{roy2008collective,yakovenko2007theory}. Here, we have identified $q_0=-\omega$. Using the definition of the electric field $\mathcal{E}_\beta(q,\omega)=i\omega b_\beta(q,\omega)-i q_\beta b_0(q,\omega)$, this term leads to a Hall contribution to the current
from Eq.~\ref{eq:linresp} given by $j_{H,\alpha}=-C_5 \epsilon_{z\alpha\beta} \mathcal{E}_\beta=C_5(\mathcal{E}\times\hat{z})_\alpha$. The role of the difference $(C_4-C_5)$ for a superconductor will be a central topic in this work. The term proportional to $C_3$ produces a term in the action $\nabla\theta\cdot\nabla \partial_t\theta=(1/2)\partial_t [(\nabla\theta)^2]$ which vanishes from being a total derivative. Therefore, we can set $C_3=0$. 

\section{Electromagnetic response}\label{sec:response}
To understand the physical implication of the coefficients $C_j$, let us compute the ac electromagnetic response as a function of 
frequency $\omega$ and wave-vector $q$. Since, we are considering rotationally symmetric systems (for simplicity) we will assume $q$ to 
be along the $x$ direction, which we will also call $L$ (for longitudinal or curl free). Since we are considering two dimensional systems,  we choose the 
other spatial direction $y$ to be perpendicular and also called $T$ (for transverse or divergence free).
Thus, $L,T$ together with $0$ for time will be the values of the indices $\alpha$ and $\beta$ in the above equations.
In this notation, the gauge-invariant electric-fields that are derived from the generalized vector potential $b_\alpha$ are written as $\mathcal{E}_T(q,\omega)=i\omega b_T(q,\omega)$ and $\mathcal{E}_L(q,\omega)=i\omega b_L(q,\omega)-i q b_0(q,\omega)$. Because of gauge-invariance, the phase fluctuation drops out of the vector $\mathcal{E}_\alpha$ and is restricted to $b_0$. Choosing (for this calculation) a 
gauge where $A_0=0$ (i.e. radiation gauge), $b_0=i\omega\phi$ represents the phase fluctuations. Applying charge conservation $(\omega j_0-q j_L)=0$ to the linear response relation Eq.~\ref{eq:linresp}  determines the phase fluctuation $b_0=\frac{q}{C_2 q^2-C_1\omega^2}[iC_2\mathcal{E}_L+(C_4-C_5)\omega\mathcal{E}_T]$. 
Substituting $b_0$ in the linear response equation Eq.~\ref{eq:linresp} leads to the ac conductivity tensor $\sigma_{\alpha\beta}(q,\omega)$ for the superconductor.
 The longitudinal conductivity tensor produces the well-known result~\cite{schrieffer2018theory} $\sigma_{LL}=\frac{iC_1 C_2 \omega}{C_2 q^2-C_1 \omega^2}$, which has a pole associated with the Goldstone phase mode. Similarly, the transverse response to lowest order in $q,\omega$, takes the standard form $\sigma_{TT}=-i C_2/\omega$, which leads to the Meissner screening response $j_T=-C_2 A_T$~\cite{altland2010condensed}. In the weak pairing limit $\Delta\ll E_F$, 
 these conductivities are unchanged from the normal state in the extreme limits $q\ll \omega$ and $q\gg \omega$. In the former case, $\sigma_{LL}=\sigma_{TT}=-i C_2/\omega$ is simply the inertial response of the electron gas that leads to 
 the plasmons. The latter case is the static Thomas-Fermi response, matches the normal response only in the longitudinal case where $\sigma_{LL}\sim i C_1 \omega/q^2$. While this may appear unfamiliar at first, the corresponding charge compressibility $\chi=(q^2/i\omega)\sigma_{LL}=C_1$ allows us to associate $C_1$ with the charge compressibility for the normal state.

Let us now consider the ac Hall response of such as superconductor~\cite{roy2008collective,Lutchyn2008gauge} arising from $C_{4,5}\neq 0$, which turns out to be 
\begin{align}\label{eq:sigmaoff}
    &\sigma_{LT}=-\sigma_{TL}=\frac{C_2 C_4 q^2-C_1 C_5\omega^2}{C_1\omega^2-C_2 q^2}.
\end{align}
While the applied electric field in the dc limit is expected to be screened, a central indicator of chirality of a superconductor is the $q\ll\omega$ ac Hall response~\cite{roy2008collective, Lutchyn2008gauge}, which in our case determines 
the coefficient $C_5$: 
\begin{align}\label{eq:acHall}
    &\sigma_{H}=\sigma_{LT}(\omega\gg q)= C_5.
\end{align}
As an aside, it was realized that the chiral nature of the superconductor does not 
contribute to $C_5$ in the translationally invariant case~\cite{Lutchyn2008gauge}, 
though it reappears in multiband superconductors~\cite{brydon2019loop}.
Explicit computation of the effective action in Eq.~\ref{eq:Seff}, similar to the 
case of the normal state, shows that the dominant contribution to $C_5$ arises from 
high energy inter-band matrix elements that are relatively unaffected by correlation 
and superconductivity. Therefore, we expect $C_5$ and the ac Hall conductivity for $q\ll\omega$ to retain the normal state anomalous Hall value, which is determined by the Berry curvature of the bands~\cite{haldane2004berry}.

Let us now consider the other limit i.e. $q\gg\omega$, which is the finite q static limit. This limit can be understood by combining the conservation relation $j_0=-q j_L/\omega=-q\sigma_{LT}\mathcal{E}_T/\omega$  with Faraday's law $\omega B_T=-q\mathcal{E}_T$, as the charge response to a flux lattice  
\begin{align}
    j_0=\sigma_{LT}(q\gg \omega)B_T=C_4 B_T\label{eq:Streda}
\end{align}
, where $B_T$ is the amplitude of the magnetic field variation in the flux lattice 
with period $q$. Physically, the modulation of the charge density $j_0$ can be viewed as the accumulation of charge in response to the application of a magnetic field. Thus, $C_4$ is the Streda response coefficient~\cite{streda1982theory}, which is 
 proportional to the Hall conductivity $\sigma_H$ in non-interacting systems~\cite{haldane2004berry}. Since $\sigma_{LT}$  arises from inter-band transitions that have a smooth frequency dependence near $\omega\sim 0$, one expects the coefficient $C_4$ to match the normal state value. For non-interacting systems one expects $C_5=C_4$, with both being related to Berry curvature~\cite{haldane2004berry}. However, for a flux lattice applied to a normal metal, a large N or RPA calculation would lead to a screening of the charge $j_0$ by a factor related to $C_1$. This would lead to a difference between $C_4$ and $C_5$.

\section{Vortex charge in an anomalous Hall superconductor}
The flux lattice discussed in the previous paragraph leads to a supercurrent 
pattern from the Meissner effect, which resembles a lattice of vortex-antivortex pairs. This motivates the question of whether a vortex, even in the absence 
of an external magnetic field, would carry a vortex charge. 
To understand the vortex charge on a lattice, let us note that a phase vortex can be converted into an anti-vortex by a large gauge transformation 
\begin{align}
    &\phi(r)\rightarrow \phi(r)+\Lambda(r)\nonumber\\\
    &\bm A(r)\cdot\delta r\rightarrow \bm A(r)\cdot\delta r+\Lambda(r+\delta r/2)-\Lambda(r-\delta r/2),
\end{align}
where $\Lambda(r)$ is a smooth function that winds by $4\pi$ 
around the center of the vortex. Note that the $4\pi$ transformation corresponds to a full electron flux quantum (as opposed to a superconducting flux quantum). On a lattice, the 
magnetic field associated with this vector potential vanishes everywhere, except for a flux quantum in one plaquette of the lattice. 
Ignoring, for the moment, the limitations of applying $S_{eff}$ to a point flux, 
the charge difference between a vortex and anti-vortex can be obtained from the Streda formula Eq.~\ref{eq:Streda} to be:
\begin{align}
&\Delta Q_v=C_4\int dr B_z=2C_4 \Phi_0,\label{eq:Qv}
\end{align}
where $\Phi_0$ is the superconducting flux quantum. This suggests  a 
charge difference between vortices and antivortices related to the Hall response 
$C_4$ as has previously been conjectured~\cite{goryo2000vortex}.

\begin{figure}
    \centering
    \includegraphics[width=\columnwidth]{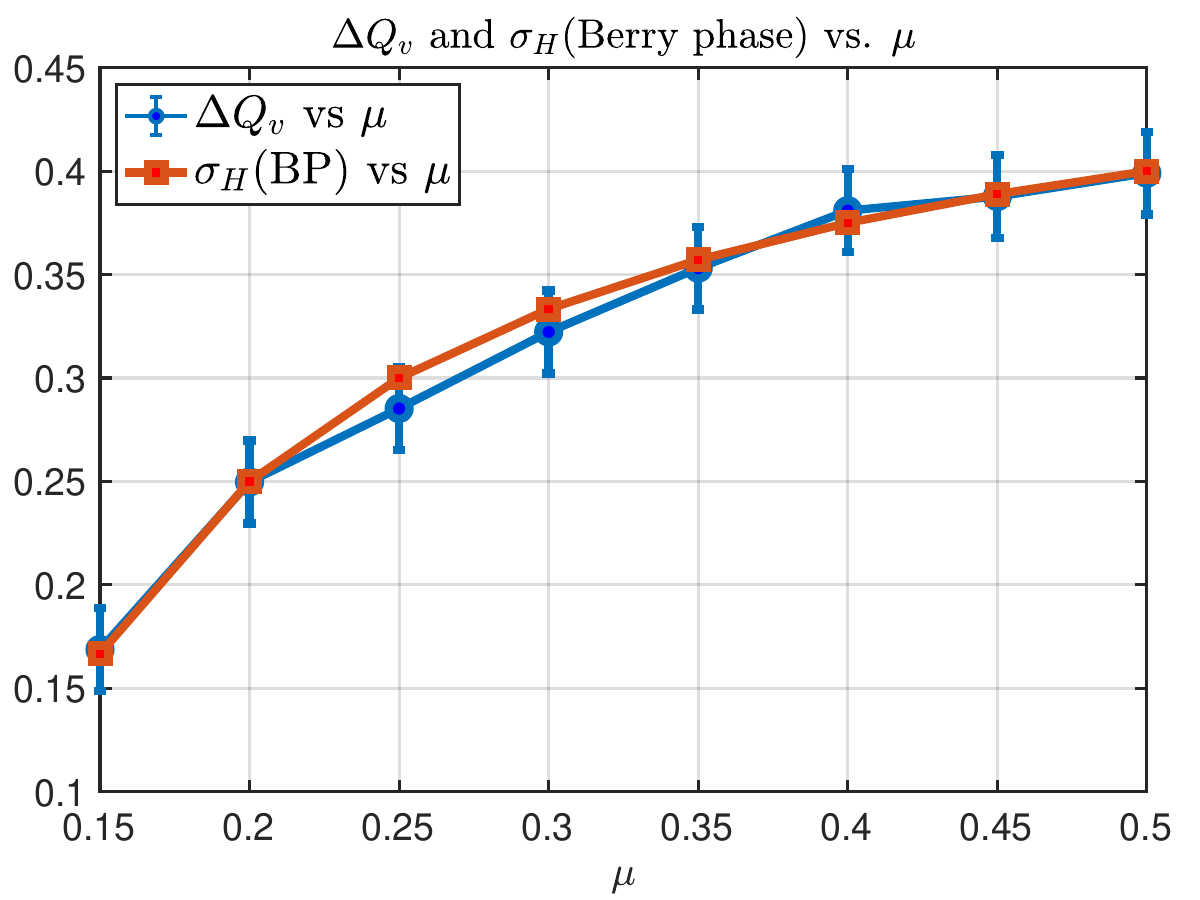}
    \caption{The charge difference between the vortex and anti-vortex, defined as $\Delta Q_v = Q_- - Q_+$, and the Hall conductivity $\sigma_H$ related to the Berry Phase are shown as functions of $\mu$ over the range $\mu = 0.15$ to $0.5$. The $\Delta Q_v$ data is presented with error bars indicating computational uncertainty. The lattice model features a size of $N = 200$, with parameters $m_0 = 0.1$, $\Delta_0 = 0.05$, and coherence length $\xi = 12.0$.}
    \label{fig:DeltaQv_SigmaH}
\end{figure}
The subtlety of applying Eq.~\ref{eq:Seff} to a point flux motivates us to numerically study the above suggestive relationship between vortex charge and the Berry phase. For this purpose we employ a model based on a bilayer gapped Dirac model on a square lattice that generates Chern number in a way similar to multilayer graphene and combine this with $p_x + ip_y$ superconducting pairing. Specifically, we consider a variation of the two-dimensional Bernevig-Hughes-Zhang (BHZ) model~\cite{bernevig2006quantum}, expressed as:
\begin{align}{\label{H0}}
H_0({\bf k}) &= \sin(k_x)\sigma_x + \sin(k_y)\sigma_y \nonumber\\
&\quad + (2 + m_0 + \cos(k_x) + \cos(k_y) - \mu)\sigma_z,
\end{align}
where the operators $\sigma_i$ represent the layer degree of freedom instead of spin. To account for the superconducting pairing, we construct the Bogoliubov-de Gennes (BdG) Hamiltonian:
\begin{align}\label{Hbdg}
H_{BdG} &= H_0 \tau_z + \Delta \sigma_{z=1} \tau_+ + h.c.,
\end{align}
where $\tau_i$ denotes the Nambu space, $\sigma_{z=1}$ indicates that the superconducting pairing is applied exclusively to the top layer ($\sigma_z = 1$), and $\Delta = \Delta_0(k_x + ik_y)$ signifies that the pairing is of the $p_x + ip_y$ type.
We introduce an (anti-)vortex into the system, we can replace $\Delta$ with its anti-commutator with the (anti-)vortex operator:
\begin{align}\label{deltavortex}
\Delta &\rightarrow \left\{\Delta, \hat{V}({\bf{r}})\right\} = \Delta_0 \left\{\hat{k}_x + i\hat{k}_y, e^{\pm i\theta_r} h(r)\right\},
\end{align}
where $\theta_r$ and $h(r)$ represent the phase and amplitude of the superconducting order parameter, respectively. The $+$ sign corresponds to a vortex, while the $-$ sign corresponds to an anti-vortex. Within the (anti-)vortex core, we have $h(r) \sim (1 - e^{-r/\xi})$, with $\xi$ being the coherence length, and $\theta_r$ possesses a winding number of $\pm 1$ around the core.

For the numerical computation of vortex charge in $H_{BdG}$, we utilize a lattice model with a size of $N=200$ under periodic boundary conditions, constructing a phase profile $\theta_r$ that features a vortex at the center and an anti-vortex at the corner. To isolate the vortex, we tweak the phase profile, ensuring that the phase around the center closely resembles an ideal isotropic vortex. We then determine the eigenstates and calculate the total charge around the vortex, denoted as $Q_+$. A similar procedure is applied to the anti-vortex to obtain its charge, $Q_-$. The charge difference between the vortex and the anti-vortex is then defined as $\Delta Q_v = Q_- - Q_+$. This analysis is performed at various chemical potentials $\mu$ and compared with the normal state Hall conductivity 
\begin{align}\label{eq:HallBerry}
    \sigma_H=\frac{1}{2}\left(1-\frac{m_0}{\mu}\right)
\end{align}
, where $m_0$ is the Dirac mass of Eq.~\ref{H0}. The result of the vortex charge versus chemical potential $\mu$ shown  in Fig. \ref{fig:DeltaQv_SigmaH}~\footnote{To verify the convergence of the numerical results, I increased the energy cutoff for each value of $\mu$ to ensure that the results converge at $N=200$. I also expanded the system size to approximately $300$, observing minimal changes (around $0.01$). Consequently, the error bars used are based on this observation.
Furthermore, I varied the coherence length $\xi$ to approximately $1.0$ and $30.0$, finding that the results remained largely unchanged.
} confirm the expectation that an anomalous Hall 
superconductor shows that the difference in vortex and antivortex charge $\Delta Q_v$ creates a charge density response that is essentially unchanged from the Streda-type formula applied to the 
anomalous Hall metal~\cite{streda1982theory,haldane2004berry}. The Streda-type 
response from vortex charges was discussed for chiral $p-$wave superconductors~\cite{Ariad2018how}.

\section{Hall response of the BKT phase:}
The vortex charge $\Delta Q_v$ plays a crucial role of describing the Hall effect in the non-superconducting phase at temperatures above the BKT transition. Specifically, let us consider a situation where $T_{BKT}$, which is controlled by the superfluid stiffness $C_1$, 
is smaller than the pairing amplitude $\Delta$ so that for a temperature $T_{BKT}\ll T\ll \Delta$ the system will be a resistive metal 
that is described by the action Eq.~\ref{eq:SE}. Such a phase can be described as being in the plasma phase of a Coulomb gas of vortex-antivortex pairs~\cite{minnhagen1987two}. The response properties of the Coulomb gas such as the resistivity and Nernst effect can be understood in terms of a duality transformation~\cite{minnhagen1987two} where the supercurrent in the superconductor 
$j=\rho_0 (\hat{z}\times \tilde{E}_v)$ maps to an electric field $\tilde{E}_v$ seen by the vortices  and the electric field $E=\Phi_0(\hat{z}\times j_v)$ is given by the vortex current $j_v$~\cite{raghu2008vortex}. 
Here $\rho_0$ and $\Phi_0$ are the superfluid density and flux quantum respectively. 
As shown in Fig.~\ref{fig:schematic}, the vortex electric field $\tilde{E}_v$,
encodes the effective Lorentz force or the Magnus force imparted to vortices by a supercurrent~\cite{minnhagen1987two}.  The vortex current $j_v$ is equivalent to a rate of phase slip generation that leads to a voltage gradient. Both the normal state conductivity and the Nernst effect can be understood from applying these duality relations to the diffusive motion of vortices~\cite{mineev2007broken,raghu2008vortex}. 
In the case of a difference $\Delta Q_v$ between vortices and anti-vortices, the vortex current $j_v$ also contributes to the total current so that we must modify the current relation as 
\begin{align}
j=\rho_0 (\hat{z}\times \tilde{E}_v)+j_v\Delta Q_v/2.
\end{align}
Assuming a diffusive vortex conductivity $\tilde{E_v}=\sigma_v^{-1} j_v=\Phi_0^{-1}\sigma_v^{-1}(\hat{z}\times E)$ leads to the relation 
\begin{align}\label{eq:HallBKT}
j=\rho_0\Phi_0^{-1}\sigma_v^{-1} E+\Delta Q_v (\hat{z}\times E)/2\Phi_0.
\end{align}
The first term is the usual Ohmic conductance in a mixed phase superconductor from flux flow~\cite{bardeen1965theory}, while the latter term is a Hall conductivity $\sigma_H$ that is clearly universally related to the vortex charge difference $\Delta Q_v$. Combining with Eq.~\ref{eq:QV}, this predicts a dc Hall response $\sigma_H=C_4$ that appears to differ from the ac Hall response $C_5$.

\section{Vortex charge screening}
The coefficients $C_{4}$ and $C_5$ that appear in the Streda-type response (i.e. Eq.~\ref{eq:Qv}) and the Hall response Eq.~\ref{eq:acHall} are, in principle, 
different. In fact, these coefficients are different even in the normal state, which serves to determine the value of $C_{4,5}$ at weak pairing. However, 
for the weakly interacting limit that we use in our numerical simulations these
coefficients are both given by the Berry curvature according to Eq.~\ref{eq:HallBerry}. Including interactions renormalizes $C_{4,5}$ differently 
as can be checked by straight-forward calculation in the large $N$ limit or using 
the random phase approximation~\cite{altland2010condensed}. This can be understood easily from Eq.~\ref{eq:Qv}, since the coefficient  $C_4$ of the Streda-type charge response should be subject to screening from interactions. The ac Hall response coefficient $C_5$ is not associated with any charge build-up and should not be screened. In fact, since the coefficient $C_5$ is related to interband transitions, 
one can relate it to the occupation function of the fermions, which would be unaffected by weak interactions. However, Eq.~\ref{eq:HallBKT} for the Hall conductivity in the BKT phase seemed to depend strongly on the vortex charge $C_4$. This leads to an apparent paradox for whether the Hall conductivity in the BKT phase is closer to the normal state value (as was suggested for superconductors in magnetic fields~\cite{bardeen1965theory}) or is renormalized.

To answer this question, we need to consider carefully the screening process of the vortex charge when a vortex-antivortex pair is formed. Studying vortex formation systematically is beyond the validity of the formalism in this work. On the other hand, the numerical results in Fig.~\ref{fig:DeltaQv_SigmaH} suggest that the vortex charge difference is quite similar to a magnetic flux, whose dynamics can be studied using the effective action in Eq.~\ref{eq:SE}. Therefore, we consider the charge response of a flux-antiflux pair, which is represented by an 
external magnetic field with a Fourier transform $B(q,t)=2 i e^{-q^2 R^2}\sin{(v q_x t)}\Theta(t)$, where $\Theta(t)$ is the Heavisider step function. This external magnetic field corresponds to a pair of fluxes with radius $R$ moving in opposite directions with velocity $v$ along $x$. The corresponding electric field from Faraday's law is transverse and written in momentum and frequency space as 
\begin{align}
    E_T=\frac{2 v q_x e^{-q^2 R^2}}{i\omega q[\omega^2-(v q_x)^2]}.
\end{align}
Using $\sigma_{LT}$ from Eq.~\ref{eq:sigmaoff} we find that the longitudinal current density $j_L$, in addition to the usual ac Hall (i.e. $\omega\gg q$) part $\j_{L,0}=C_5 E_T$ contains an additional "screening" contribution, which is proportional to $C_4-C_5$: 
\begin{align}
\delta j_{L}=\frac{(C_4 -C_5)c^2 q}{\omega^2-c^2 q^2}\frac{2i \omega q_x v e^{-q^2 R^2}}{[\omega^2-(v q_x)^2]},
\end{align}
where $c=\sqrt{C_2/C_1}$ is the plasmon velocity. 
Fourier transforming this component to the time-domain yields: 
\begin{align}
    &\delta j_{L}=2i (C_4-C_5) v c^2 q_x q e^{-q^2 R^2}\frac{[\cos{c q t}-\cos{v q_x t}]}{c^2q^2-v^2 q_x^2}.
\end{align}
The contribution to the above proportional to $\cos{v q_x t}$ combined with the near field part (i.e. proportional to $1-e^{-q^2 R^2}$) of $E_T$ corresponds to the flow of vortex core charge density shown in Fig.~\ref{fig:schematic}
proportional to $C_4$. On the other hand, the contribution to $\delta j_L$ from $\cos{c q t}$ contributes to the crescent shaped charge waves in Fig.~\ref{fig:schematic}.
The combined result $\delta j_L$ in the above equation clearly vanishes as $q,q_x\rightarrow 0$ establishing that the longitudinal current response is determined by $j_{L,0}$, which is proportional to the high-frequency ac Hall conductivity $C_5$, despite screening reducing the charge at the vortex core to $C_4$.
The longitudinal current response in vector form is 
\begin{align}
    \vec{j}_{L,0}=C_5 (\hat{z}\times \vec{E}_T)=\vec{q}\frac{ 2 v q_x e^{-q^2 R^2}}{q^2}C_5  \cos{v q_x t}.
\end{align}
Note that while the x component of the current approaches a constant $j_{L,0,x}\sim 2 v C_5$ as $q_x=q\rightarrow 0$, the current has a non-trivial dependence on $q_y/q_x$, which reflects the angular dependence of the far field that can lead to logarithmic in system size corrections to $j_{L,0}$. This does not, however, affect the conclusion that the vortex Hall conductivity is determined by the high frequency ac Hall conductivity $C_5$.

\section{Conclusion}
We have studied the dc anomalous Hall response of a superconductor above 
the BKT transition but below the mean-field superconducting gap, where a vortex plasma phase is responsible for dissipative transport. Based on the effective action~\ref{eq:Seff}, we conjecture based on an analogy between fluxes and vortices, that the core charge of a vortex and anti-vortex might differ by an amount proportional to the Streda response coefficient $C_4$, which in non-interacting metals is expected to be determined by the Fermi surface Berry phase~\cite{haldane2004berry}. In Fig.~\ref{fig:DeltaQv_SigmaH}, we numerically verify this for a superconducting version of the BHZ model. The coefficient $C_4$, however differs in interacting fermion systems from the ac Hall conductivity $C_5$. By using the analogy between fluxes and vortices together with a flux flow model~\cite{bardeen1965theory} for superconducting transport shown in Fig.~\ref{fig:schematic} we showed that the dc Hall conductivity should actually match the ac value $C_5$.
We expect the effective action Eq.~\ref{eq:Seff} with coefficients $C_{j=1,2,4,5}$ to be a good description of any chiral superconductor including tetra-layer graphene with coefficients that are measurable in linear response. It would be interesting to compare these coefficients to  vortex charge as well as dc Hall conductivity measurements.

\begin{acknowledgments}
We thank Maissam Barkeshli, Yang-zhi Chou, Jihang Zhu and Seth Musser (for telling us about periodic vortex/antivortex phase configurations ) for valuable discussions. 
J.S.\@ acknowledges support from the Joint Quantum Institute and the hospitality of the Aspen Center for Physics, which is supported by National Science Foundation grant PHY-2210452.
This work is also supported by the Laboratory for Physical Sciences through its continuous support of the Condensed Matter Theory Center at the University of Maryland.
\end{acknowledgments}%

\bibliography{bib.bib}
\appendix
\section{A: Effective action for the superconductor}
Let us consider a simplified action for a superconductor in an anomalous Hall 
metal, which is obtained by applying a Hubbard-Stratonovich decomposition of an 
attractive interaction~\cite{altland2010condensed} and is written as:
\begin{equation}\label{eq:S0}
    S[A,\phi]=\int [\bar{\psi}(i\partial_t-A_0)\psi-\bar{\psi}h\psi+\{\psi^T\Delta\psi+c.c\}],
\end{equation}
where the action $S$ depends on the vector potential through $h[A](x,x')=exp[i\int_0^1 d\lambda A(x+\lambda(x'-x))]h(x,x')$ and to the superconducting phase $\phi$ through the relation $\Delta[\phi](x,x')=e^{i(\phi(x)+\phi(x'))/2}\Delta(x,x')$,
where $\Delta$ is the anti-symmetric in space superconducting pairing potential. 
Here, for simplicity, we have ignored spin and valley degrees of freedom. 
For the purpose of integrating out the fermions, it is convenient to introduce 
Majorana or real Grassmann's $\psi=\gamma_1+i\gamma_2$ so that the action can be 
written in a Nambu-matrix form $S[A,\phi]=\int \Gamma^T G^{-1}\Gamma$ where 
$G^{-1}=[(i\partial_t-A_0- h_a)\tau_0+h_s\tau_y+\Delta_r\tau_z+i\Delta_i\tau_x],$
is the inverse Nambu-Gorkov Green function and  $\tau_\alpha$ are the Pauli matrices in the Nambu spinor space
and $\Gamma(x)=(\gamma_1(x),\gamma_2(x))^T$ is the Majorana spinor. The phase fluctuations $\phi$ and vector potentials obey the gauge transformations $\phi\rightarrow \phi-\Lambda$ and $A_\alpha\rightarrow A_\alpha-\partial_\alpha\Lambda$. Thus, we can use a gauge transformation to eliminate the phase fluctuations in terms of gauge-invariant fields $b_\alpha=A_\alpha-\partial_\alpha \phi$. 

\section{B: Explicit form of response matrix $K$}
For a 
two dimensional gapped superconductor with rotational symmetry, at small wave-vectors and frequencies, we can expand $K^{(\alpha,\beta)}(q,\omega)$ up to linear order in $q$ and $\omega$ so that the response functions can be written as
\begin{align}
    &K^{(0,0)}(q,\omega)=-C_1,\quad K^{(\alpha,\alpha)}(q,\omega)=C_2,\nonumber\\
    & K^{(\alpha,0)}(q,\omega)=-K^{(0,\alpha)}(q,\omega)=i q_\beta (C_3\delta_{\alpha\beta}+C_4\epsilon_{z\alpha\beta }) ,\nonumber\\
    & K^{(\alpha,\beta)}(q,\omega)=-i\omega\epsilon_{z\alpha\beta }C_5,
\end{align}
where the indices $\alpha,\beta$ represent the spatial directions $x,y$. 
\end{document}


\title{Supplementary Material for ``Exact Andreev reflections from impurities in quantum Hall point contacts and Luttinger liquids''}

\author{Stuart N. Thomas}
\author{Jay D. Sau}

\date{June 25, 2024}
\maketitle

\section{Numerical Details}
The $K=1/3$ Luttinger liquid poses a challenge to simulate numerically since there is no explicit form for the parameters in Eq.~\ref*{eq:numerical-ham} of the main text which lead to such a scaling coefficient. To calculate the corresponding parameters numerically, we measure the Drude weight and the compressibility and relate them back to the Luttinger parameters as $D=Kv/2\pi$ and $\chi=K/\pi v$. We use the VUMPS algorithm \cite{vumps} to determine these quantities for infinite systems to remove finite-size effects, particularly in calculating the compressibility. The Luttinger parameter is first set by optimizing the ground state $\psi$ over $K^2(\psi)=2\pi^2 D(\psi) \chi(\psi)$. Then, the velocity is tuned by rescaling $t_n$ and $\mu_n$. We use a large field to avoid the gapped phase near $\mu_n=0$ when $U_n>0$. We find that 
$t_n=0.122$, 
$U=0.293$, 
$U'_n=0.439$ and 
$\mu_n=1.22$ 
gives parameters of $v=0.5$\footnote{We choose $v=0.5$ instead of $v=1$ to access smaller frequencies without increasing the wavelength.} and $K=0.332$. To match the magnetization and velocity on the $K=1$ side, we set the hopping to 
$t_n = 0.308$ and the chemical potential to 
$\mu_n =(4t_n^2-v^2)^{1/2}= 0.360$.

We represent the interface between these two phases as a slow variation of $K(x)$ relative to the lattice spacing, which corresponds to slow variations in the lattice parameters. The Hamiltonian in Eq.~\ref*{eq:numerical-ham} of the main text can be transformed by a Jordan-Wigner transformation to an XXZ Hamiltonian. 

Introducing smooth variations of $\mu_n$ and $a_n$ can be used to generate wavepackets of chiral charge to numerically simulate scattering in the lattice fermion model~\cite{vlijm2015quasi,ganahl2012observation}. To understand this, we first apply a gauge transformation (i.e. time-dependent unitary transformation) to set $\mu_n=0$ and transform $\tilde{a}_n(t)=a_n-\int_{-\infty}^t d\tau (\mu_{n+1}-\mu_n)$. In the continuum limit where $\tilde{a}_n$ is slowly varying, the field is Lorentz invariant so that a variation $\tilde{a}_n=f(v t-n)$ is Lorentz invariant. In fact, such a vector potential represents a wavepacket of chiral charge 
that propagates with velocity $v$. We can apply a gauge transformation so that $\partial_t a_n|_{t=0}=0$ and we get $\mu_n=\sum_{m<n}f'(t-m)$ and $a_n=f(t-n)$.
 In our simulations, we use a Gaussian for $f(x)$. To generate Figs.~\ref{fig:trajectory} and \ref{fig:ac_conductance} of the main text, we run simulations on a 640 site system with a 3 site crossover region. We use a Trotter step size of $\tau=2$ which we find is sufficient due to the slow dynamics.

\section{Refermionization details}
To formalize the refermionization procedure in accordance with Ref.~\onlinecite{von1998bosonization}, we consider the fields $\xi(x)$ and $\eta(x)$ on the finite wire $x \in (0,L)$. In this case, we can expand each as a Fourier series
\begin{align*}
\xi(x) &= A_0 + \sqrt{2} \sum_{n=1}A_n\, e^{\pi n a/L} \; \cos \frac { \pi n x} L \\
\eta(x) &= B_0 + \sqrt{2}\sum_{n=1}B_n\, e^{\pi n a/L} \; \cos \frac { \pi n x} L 
\end{align*}
where only cosine terms are included since $x>0$ and where $a$ is a soft UV cutoff for the coefficients $A_n$ and $B_n$.
From the canonical commutation relation $[\xi(x),\eta(x')]=i\pi \delta(x-x')$, we find
\begin{align*}
    [A_m, B_n] &= \frac {\pi i}{L} \; e^{-2\pi n a/L} \; \delta_{mn} \\
(a\rightarrow 0) \qquad &= \frac {\pi i}{L} \; \delta_{mn}.
\end{align*}
The chiral field $\chiral\xi(x)=(1/2)[\xi(|x|)+\int_0^x\mathrm dx\, \eta(|x|)]$ becomes 
\begin{equation*}
\chiral\xi(x) = \frac{A_0}{2} + \frac{B_0 x}{2} + \frac 1 {\sqrt{2}} \sum_{n=1} e^{-\pi n a/L}\; \left[A_n \cos \frac { \pi n x} L + \frac{B_n L}{\pi n} \sin \frac { \pi n x} L\right]
\end{equation*}
which obeys 
\begin{align*}
[\chiral\xi(x),\chiral\xi(x')] &= -\frac i 2 \sum_{n=-\infty}^\infty \tan^{-1} \left( \frac{x-x'-2nL} a\right) \\
(a\rightarrow 0) \qquad &= -\frac{i\pi}{4} \, \mathrm{sgn}(x-x').
\end{align*}
This relation lacks the $O(1/L)$ term found in Ref.~\onlinecite{von1998bosonization} due to the inclusion of the zero modes $A_0$ and $B_0$. These zero modes become the Klein factor and $\exp(2\pi N x/L)$ factors in the bosonization identity of Ref.~\onlinecite{von1998bosonization}.

The Hamiltonian in Eq.~\ref*{eq:ham2} of the main text then becomes
\begin{equation}
H_\xi = \int_0^L \frac{\mathrm dx}{\pi} \, (\partial_x \chiral\xi(x))^2 + g\cos 2\chiral\xi(0).
\end{equation}
In the noninteracting limit ($g=0$), the equation of motion becomes $\partial_t \chiral\xi=\partial_x \chiral\xi$, demonstrating that $\chiral\xi(x)$ is in fact chiral.

One subtlety in this refermionizion procedure is that $\xi(x)$ and $\eta(x)$ are not well-defined operators on the Hilbert space of states. 
This concern was mentioned in Ref.~\onlinecite{von1998bosonization} and remedied with explicit Klein factors. 
Consider for example the unitary transformation $T=\mathrm{exp}(2\pi i N)$ where $N = \pi^{-1} \int_0^L \mathrm dx\, \eta(x)=B_0 L/\pi$ is the chiral fermion number. 
Since $[N,\xi(x)]=i$, the transformed field is $T \xi(x) T^\dagger=\xi(x)+2\pi$ which leads to the equivalence $A_0 \equiv A_0 + 2\pi$. 
In essence, the operator $A_0$ is periodic and only defined on a compact subspace. Since our refermionization prescription (Eq.~\ref*{eq:refermionization} of the main text) only depends on $A_0$ only up to this transformation, the fermion field $\chiral\psi$ is well-defined. In other words, we do not utilize expressions of the form $\mathrm{exp}(ic \chiral\xi(x))$ where $c\notin \mathbb Z$.

\section{Andreev Reflection}
The transmission due to the interaction term in Eq.~\ref*{eq:cosF} on the main text can be solved by solving the time evolution of $\gamma_1(x)$ and $\alpha$ using the Heisenberg equation of motion in the frequency domain:
\begin{align*}
\omega \gamma_1(x) &= -[H_{\gamma_1}, \gamma_1(x)] \\
&= i \partial_x \gamma_1(x) - 2\sqrt{2\EB} \alpha \delta(x), \\
\omega \alpha &= -[H_{\gamma_1}, \alpha] \\
&= 2\sqrt{2\EB} \gamma_1(x).
\end{align*}
Integrating the region around $x=0$ gives 
\begin{align*}
\gamma_1(0_+) - \gamma_1(0_-) &= i \sqrt{2\EB} \alpha \\ 
\gamma_1(0_+) + \gamma_1(0_-) &= \frac{2\omega }{\sqrt{2\EB}} \alpha.
\end{align*}
The transmission coefficient is
\begin{align*}
T &= \gamma_1(0_+) / \gamma_1(0_-) \\ 
&= \frac{\omega + i\EB}{\omega-i\EB}
\end{align*}
which is a pure phase. Transforming back to the fermion basis, we calculate the transmission matrix of $\chiral\psi$ to be
\begin{equation*}
t_0 = \frac 1 {iE + \EB} \left( \begin{matrix} -\EB & iE \\ iE & -\EB
\end{matrix}\right).
\end{equation*}
Refolding this system using Eq.~\ref*{eq:dirac-fermion} of the main text, this transmission matrix leads to a reflection matrix of 
\begin{align*}
r &=\sigma_x t_0 \\
&= \frac 1 {iE + \EB} \left( \begin{matrix} iE & -\EB \\ -\EB & iE
\end{matrix}\right)
\end{align*}
which leads to the amplitudes in Eq.~\ref*{eq:rAndreev} of the main text.

\section{Review of Noise}
Using Eq.~\ref{eq:rAndreev} we can reproduce known formulae for the conductance and shot noise \cite{chamon1997distinct,sandler1998andreev,sandler1999noise}, the latter of which reveals Andreev reflection in the form of the charge of the scattered observables~\cite{muzykantskii1994quantum,de1994doubled}. 
 The noise at high voltages $V \gg \EB$ goes as $P\propto \EB\ll I\sim V/2$ resulting from incoherent normal reflection caused by the impurity that would vanish 
 in the absence of backscattering. Furthermore, the noise power nearly vanishes relative to the current as a result of most of the current being carried 
 by perfect Andreev reflection. In this case, the noise power approaches $P\sim \pi e^2\EB/2 h \sim \delta I$
 where $\delta I$ is the small back-scattered component of the current at high voltages. The effective charge of the back-scattered current is $P/2\delta I\sim e/2$.
 On the other hand at low voltages the power goes as $P\sim V^3/12 \EB^2\sim 2 I$, which is consistent with tunneling of electrons 
 as expected from the fraction at $\nu=1$. This is because the effective charge $P/2\delta I \sim e$, which is twice the high voltage value. This is similar to tunneling in superconductors where the high voltage noise from quasiparticles is twice the result from Cooper pair tunneling at low voltages. 
 
The asymptotic behavior of the shot noise is plotted in Fig.~\ref{fig:noise}.
\begin{figure}[h]
\centering
\includegraphics[width=0.6\columnwidth]{noise_current.pdf}
\caption{\label{fig:noise} Asymptotic behavior of noise $P$ representing $V\rightarrow \infty$ (top) and $V\rightarrow 0$ (bottom) asymptotic behavior. The $T=0$ case matches $P\rightarrow \delta I$ in the large voltage limit and $P\rightarrow 2 I$ in the small voltage limit. The impurity energy is $\EB=1$.}
\end{figure}

\section{Model for QPC with a Quantum Hall transformer}
Here we discuss a possible model for a QPC that is wider than the magnetic length 
so that there is a segment of domain-wall across the QPC. The QPC has 5 segments of chiral fractional quantum Hall edges. The 
density $\rho(s,t)$ on each of the edges, which are the only compressible regions in the QPC, is written as 
\begin{align}
    &\partial_t\rho-\partial_s [v(s)\rho(s)]=\nu \partial_s \phi(s),
\end{align}
where $s$ is a coordinate along the relevant edge segment, $v(s)$ is the edge velocity, $\nu$ is the effective filling factor of the 
edge and $\phi(s)$ is the electrostatic potential of the device. 
The edge current is $v(s)\rho(s)$, while the RHS represents an anomaly contribution from the bulk quantum Hall conductance.
For a steady state system, as is relevant for computing conductivity, this implies 
\begin{align}
    v_a(s)\rho_a(s)+\nu_a(s) \phi_a(s)=\mu_a(s) \nu_a(s)=\textrm{constant}\label{eq:rho}
\end{align}
over a certain segment of edge labelled by $a$. For the four segments of physical edges of the QPC, the label $a=(\mathrm t/\mathrm b,\pm)$ corresponding to the top and bottom edges of the QPC in Fig.~\ref{fig:domain_wall} (main text) and $\pm$ referring to the sign of the coordinate $s$ with $s=0$ referring to the position of the domain-wall. In this notation,  $\nu_a=\nu_\pm$ depending on filling fraction of the adjacent bulk for $s>0$ or $s<0$.
Referring to Fig.~\ref{fig:domain_wall} and keeping in mind current conservation the chemical potentials $\mu_a$ for the four
edges of the QPC are $\mu_a=\mu_{\pm}\pm j/2\nu_{\pm}$.
For the dotted domain-wall the effective filling factor is $\nu=\nu_--\nu_+$, 
though this value only represents the charged mode; the neutral mode~\cite{kane1994randomness} is not shown in the diagram. The value of the constant on each edge depends on the chemical potential 
at each edge which is determined by local current conservation at the junctions~\cite{chang2003chiral}. The electrostatic potential $\phi( r)$
is determined from the total charge density $\rho(r)$ through 
\begin{align}
    &\phi(r)=\phi_\mathrm{ext}(r)+\int \mathrm d^2r'\, U(r-r')\rho(r'),
\end{align}
where the two dimensional coordinate $r$ includes both the bulk and edges of the QPC as well the domain-wall.

For simplicity, we will consider the limit where the domain-wall edge mode velocity and string  tension $T_\mathrm{S}$ is large enough to make the domain-wall density of states and displacement from the narrowest point small. 
This assumption makes 
the domain-wall contribution to the total charge negligible and is equivalent to the assumption about edge dominance made 
in the main text. The above equation for the electrostatic potential can now 
be decomposed into edge contributions
\begin{align}
    &\phi_a(s)=\phi_{\mathrm{ext},a}(s)+\int \mathrm d s'\, \sum_b U_{ab}(s,s')\rho_b(s').\label{eq:Coul0}
\end{align}
Using Eq.~\ref{eq:rho} this equation can be rewritten as
\begin{align}
    &\int \mathrm d s'\, \sum_b \left[U_{ab}(s,s')+\frac{\delta(s-s')\,\delta_{ab}v_a(s)}{\nu_{a}(s)}\right]\rho_b(s')=\mu_a-\phi_{\mathrm{ext},a}(s),\label{eq:Coul}
\end{align}
which is a well-defined inversion of a positive-definite symmetric matrix for $\rho_b(s)$. 

The negligible bias dependence of the density near the QPC implies that the electric field cannot contribute momentum change near the 
domain wall. 
This implies that the change in canonical momentum of the edge carriers resulting from charge leaving and entering the 
top and bottom ends of the domain-wall with vector potential $A_a=\pm B W/2$ must be generated by the interaction of the domain wall end density with the 
electric field at the domain-wall. 
Because the dominant ground state contribution to the momentum 
is from the vector potential,  the rate of change of canonical momentum (i.e. effective force) amounts to the difference of currents entering the top and bottom edge that must then balance against the electrostatic force, i.e. 
\begin{multline}
\left(\mu_-+\frac j {2\nu_-}-\phi_\mathrm{t}(s=0)\right)\nu_- - \left(\mu_++\frac j {2\nu_+}-\phi_\mathrm{t}(s=0)\right)\nu_+ \\
 = \left(\mu_+-\frac j {2\nu_+}-\phi_\mathrm{b}(s=0)\right)\nu_+ - \left(\mu_--\frac j {2\nu_-}-\phi_\mathrm{b}(s=0)\right)\nu_-+\sum_{a=\mathrm t,\mathrm b} Q_a \phi'_{a}(s=0),
\end{multline}
where we have used the chemical potential notation of Fig.~\ref{fig:domain_wall} together with Eq.~\ref{eq:rho} and $Q_{\mathrm t,\mathrm b}$ are the charges at the top and bottom ends of the domain-wall. The terms $\phi_{\mathrm t,\mathrm b}(s=0)$ are 
the electrostatic potentials at the top and bottom ends of the domain wall in Fig.~\ref{fig:domain_wall}.
The average of the top and bottom potentials are:
\begin{align}
    \frac{\phi_\mathrm{t}(s=0)+\phi_\mathrm{b}(s=0)}{2}=\frac{\mu_+\nu_+-\mu_-\nu_-}{\nu_+-\nu_-}+ \sum_{a=\mathrm t,\mathrm b}  \frac{Q_a \phi'_{a}(s=0)}{\nu_+-\nu_-}.\label{eq:Pcons}
\end{align}

The above form motivates us to consider the potential and density in Eq.~\ref{eq:Coul} in a rotated basis where the indices $a,b$ in both 
Eq.~\ref{eq:Coul} and Eq.~\ref{eq:rho} refer to symmetric and antisymmetric combinations of the top and bottom edges (i.e. $a,b=S$ and $a,b=A$ respectively). 
Note that because 
the Luttinger parameter is the same of the top and bottom edges, this does not modify Eq.~\ref{eq:rho}; however the Coulomb matrix $U$ is
rotated from the $t,b$ basis to the $S,A$ basis. The effective chemical potential after this rotation transforms to $\mu_\pm$ in the symmetric sector and  $j/2\nu_\pm$ in the antisymmetric sector. 
Finally, we note that as a matter of principle, Eq.~\ref{eq:Coul} can be solved to obtain both $\phi_{\mathrm{S,A}}(s=0)$ and $\phi_{\mathrm{S,A}}'(s=0)$  as a function of $\mu_{\pm}$ and $j$. Substituting into Eq.~\ref{eq:Pcons} leads to a linear equation in these variables with coefficients $Q_{\mathrm t, \mathrm b}$, which are the charges of the domain-wall of the form
\begin{align}
    \left(C_++\sum_a D_{+,a}Q_a\right)\mu_++\left(C_-+\sum_a D_{-,a}Q_a\right)\mu_-+\left(C_j+\sum_a D_{j,a}Q_a\right)j=\sum_a \int \mathrm ds\, F_a(s)\phi_{\mathrm{ext},a}(s),
\end{align}
where $C_a$, $D_{c,a}$ and $F_a(s)$ are functions determined in terms of $U_{ab}(s,s')$ and $\nu_a(s)$.
Note that the RHS of the equation does not depend on $\mu_{\pm}, j$ and  would a drop out of a linear response calculation.
Focusing on linear response, this constraint reduces to 
\begin{align}
    \left(C_++\sum_a D_{+,a}Q_a\right)\delta\mu_++\left(C_-+\sum_a D_{-,a}Q_a\right)\delta\mu_-+\left(C_j+\sum_a D_{j,a}Q_a\right)\delta j=0.\label{eq:Pcons1}
\end{align}
Since we cannot (without more microscopic details) determine these charges microscopically, we 
need to impose a consistency condition that Eq.~\ref{eq:Pcons} allows an equilibrium.
This would lead to a consistent solution for the equilibrium state $\delta \mu_a=\delta\mu$ for any chemical potential with $\delta j=0$.
This leads to the constraint on the charges
\begin{equation}
    \left(C_++\sum_a D_{+,a}Q_a\right)+\left(C_-+\sum_a D_{-,a}Q_a\right)=0.
\end{equation}
With this constraint Eq.~\ref{eq:Pcons1} reduces to 
\begin{align}
    &\left(C_++\sum_a D_{+,a}Q_a\right)\left(\delta\mu_+-\delta\mu_-\right)+\left(C_j+\sum_a D_{j,a}Q_a\right)\delta j=0.\label{eq:Pcons2}
\end{align}
Based on Fig.~\ref{fig:domain_wall}, the bias voltage would be
\begin{align}
    &\delta V_b=\delta \mu_+-\delta \mu_-+\delta j\left(\frac{1}{2\nu_+}+\frac{1}{2\nu_-}\right).
\end{align}
the conductance would be 
\begin{align}
    G&=\frac{\delta j}{\delta V_b}\\
    &=\left[\frac{1}{2\nu_+}+\frac{1}{2\nu_-}-(C_j+\sum_a D_{j,a}Q_a)(C_++\sum_a D_{+,a}Q_a)^{-1}(\delta\mu_+-\delta\mu_-)^{-1}\right]^{-1}.
\end{align}

Determining the relevant coefficients $C_j, D_{j,a}$ based on Eq.~\ref{eq:rho}, Eq.~\ref{eq:Coul} and Eq.~\ref{eq:Pcons} is nontrivial. 
On the other hand, as mentioned in the main text, it is rather straightforward to solve in the case of a QPC with mirror symmetry.
In this case, the equations decouple in the symmetric and antisymmetric (i.e. $S$ and $A$) sectors with $\delta\mu_\pm$ belonging to $S$ and 
$\delta j$ belonging to $A$. Furthermore, symmetry dictates $Q_\mathrm t=Q_\mathrm b$, which ensures that Eq.~\ref{eq:Pcons} would belong in the sector $S$. 
This implies that $C_j=D_{j,a}=0$.
Substituting into the above equation for conductance we get 
\begin{equation}
    G=\left[\frac{1}{2\nu_+}+\frac{1}{2\nu_-}\right]^{-1},
\end{equation}
consistent with the result in the main text.

\section{Impurity Strength}
\newcommand\lB{\ell_{\mathrm{B}}}
We estimate the Luttinger liquid back-scattering parameter $g$ in Eq.~\ref{eq:ham} in the main text based on the configuration of the 
QPC in the experiment~\cite{cohen2023universal}.
In the open junction case, we assume that the 2D electron density interpolates smoothly between $n_{1/3}=1/2\pi\lB^2$ and $n_{1}=(1/3)/2\pi\lB^2$ over a width $L$ where $\lB=\sqrt{\hbar/eB}$ is the magnetic length.
From the Supplementary Material of \onlinecite{cohen2023universal}, this width is on the order of 100 \si{nm}. 
This gives a background charge gradient $\nabla\rho_0(x) =e/3\pi L\lB^2$. 
Generically, we expect a $\nu=2/3$ edge mode to develop between these two regimes, which we model as a correction $\delta\rho(x)\propto (x/\lB)\, \mathrm{exp}(-x^2/2\lB^2)$ to the density profile. The overall prefactor to this expression is determined by the constraint that $\nabla\rho(x)=\nabla(\rho_0(x) + \delta\rho(x))=0$, which leads to
\begin{equation}
\delta\rho(x) = -\frac{e x}{3\pi \lB^2 L} \mathrm{exp}\left[-\frac{x^2}{2\lB^2}\right].
\end{equation}

If this edge mode propagates along the $y$ direction over a width $W$, the potential at the edge ($y=\pm W/2$) is given by the Greens function of a charged wire
\begin{equation*}
G(x) = \log\left[\frac{|x|}{-W+\sqrt{W^2+x^2}}\right],
\end{equation*}
leading to
\begin{equation*}
V(x) = -\frac 1 {4\pi\epsilon} \int_{-\infty}^{\infty} \mathrm d x'\, G(x')\, \delta \rho(x-x'),
\end{equation*}
noting that the potential from $\rho_0$ is absorbed into the harmonic Luttinger liquid.

Using the results from Ref.~\onlinecite{kane1992transmission}, the bosonized backscattering parameter $g$ is proportional, at first order, to the Fourier transform of the potential at twice the Fermi momenum $2\kF$. Since the real-space potential is a convolution, the Fourier transform is just a product:
\begin{align*}
\tilde V(2\kF) &= -\frac {\sqrt{2\pi}} {4\pi \epsilon} \tilde G(2\kF) \, \delta \tilde\rho(2\kF) \\
%
&= -\frac {ie\kF\lB} {3\pi \sqrt{2\pi} L \epsilon} \, e^{-2\kF^2 \lB^2}\,  \tilde G(2\kF).
\end{align*}
We can approximate $\tilde G(2\kF)$ by taking the large $W$ limit in which $G(x)\approx \log|x|$ (up to a constant) and $\tilde G(2\kF)=-\kF^{-1} \sqrt{\pi/8}$. 
Then
\begin{equation*}
\tilde V(2\kF) = \frac {ie\lB} {12\pi L \epsilon} \, e^{-2\kF^2 \lB^2}.
\end{equation*}
This expression relates to $g$ as $g\propto a_0^{-1} e \tilde V(2\kF)$ where $a_0$ is the UV cutoff of the bosonization \cite{kane1992transmission}. 
Therefore
\begin{equation*}
g \propto \frac {e^2 \lB } {12 \pi a_0 L \epsilon} \, e^{-2\kF^2 \lB^2}.
\end{equation*}
This expression has units of energy, agreeing with the scaling dimension of the boundary sine Gordon model. 
Note that the $a_0$ used here and the $a$ in the refermionization section of the main text are not necessarily equal.

Crucially, the Fermi momentum in the quantum hall depends on the spatial separation of the chiral modes. This is due to the magnetic flux through the bulk causing a momentum different between the chiral modes. At the QPC, the shorter width $W$ leads to a lower Fermi momentum. This relationship is quantified as 
\begin{equation*}
\kF = \frac{\kappa W}{2\lB^2},
\end{equation*}
where $\kappa$ is a factor of proportionality.
This form leads $g$ to be exponentially suppressed in width:
\begin{equation*}
g \propto \frac {E_C \lB^2 } {3 a_0 L} \, e^{-\kappa^2 W^2 /2\lB^2}.
\end{equation*}
where $E_C=e^2/4\pi \epsilon \lB$ is the Coulomb energy defined in Ref.~\onlinecite{cohen2023universal}, which in the experimental setup is approximately $37~\si{meV}$ at a magnetic field of $10~\si{T}$.

In the refermionized expressions for noise and conductance, the impurity is parameterized by a barrier enegy $\EB=g^2 \pi a / \hbar v$, which becomes
\begin{align*}
\EB \propto \frac {E_C^2 a \lB^4 } {\hbar v a_0^2 L^2} \, e^{-\kappa^2 W^2 /\lB^2}
\end{align*}
(omitting the $\pi/9$ prefactor).
Comparing our results with Ref.~\onlinecite{cohen2023universal}, the barrier energy relates to $T_0$ simply as $2\EB=\kB T_0$.
The drift velocity $v$ of a skipping state is given as $v=|\nabla V|/B$.
Using the definition $E_V=e|\nabla V|\lB$ from the Ref.~\onlinecite{cohen2023universal} supplementary material, we find $v=E_V \lB / \hbar$, leading to
\begin{align*}
\EB \propto \frac {E_C^2 a \lB^3 } {E_V a_0^2 L^2} \, e^{-\kappa^2 W^2 /\lB^2}.
\end{align*}

In Ref.~\onlinecite{cohen2023universal}, the junction width is reduced by increasing the north-south gate voltage $V_\mathrm{NS}$. The edge state is approximately localized at the point where the drop in potential from the bulk value is equal to the spacing of the Landau levels, given by the cyclotron frequency in graphene \cite{yin2017}
\begin{align*}
\omegac &= \sqrt{\frac{2a e \vF^2 B} \hbar} \\
&= \sqrt{2} \, \frac \vF \lB 
\end{align*}
where $\vF=1.1\times10^6 \si{m/s}$. We model the potential profile as a Gaussian of width $W_0$ (standard deviation $W_0/2$) in the $y$ direction (North-South in the language of \onlinecite{cohen2023universal}). From Fig.~S1 we see that this value is on the order of $100\si{nm}$. The width of the Gaussian at an energy $\hbar \omegac$ is then
\begin{align*}
W &= 2W_0\sqrt{\log \left(\frac {e\VG} {e\VG-\hbar\omegac}\right)} \\
&\approx 2W_0\sqrt{\frac{\hbar\omegac} {e\VG}}
\end{align*}
Plugging this in to our expression for $\EB$ gives
\begin{align*}
\EB \propto \frac {E_C^2 a \lB^3 } {E_V a_0^2 L^2} \; \mathrm{exp}\left[-\frac{4\kappa^2 W_0^2 \hbar \omegac} {e \VG \lB^2 }\right].
\end{align*}
Using $E_V\approx e \VG \lB W_0/2$, and assuming that $W_0$ and $L$ are of similar order ($\approx 100\si{nm}$), we find
\begin{align*}
\EB \propto \frac {E_C^2 a \lB^2 } {e\VG a_0^2 W_0} \; \mathrm{exp}\left[-\frac{4 \kappa^2 W_0^2 \hbar \omegac} {e \VG \lB^2 }\right].
\end{align*}
We can approximate the cutoff $a$ as $L\sim W_0$ since our refermionized picture assumes a sharp crossover between free fermions and we take the standard approximation $a_0\sim \kF^{-1}=\lB^2/W_0$, using $W_0$ in lieu of $W$ to ensure that this cutoff is constant as we tune $\VG$. Then,
\begin{align*}
\EB \propto \frac {E_C^2 \kappa^2 W_0^2} {e\VG \lB^2} \; \mathrm{exp}\left[-\frac{4 \kappa^2 W_0^2 \hbar \omegac} {e \VG \lB^2 }\right].
\end{align*}
This expression can be simplifies by considering two parameters of the experimental system: $W_0/\lB\approx 12.3$  and $\hbar \omegac=\sqrt{2}\hbar\vF/\lB\approx 0.126 \si{eV}$ at $B=10\si{T}$.
\begin{align*}
\EB \propto \frac {E_C^2 a \lB^2 } {e\VG a_0^2 W_0} \; \mathrm{exp}\left[-\frac{4 \kappa^2 W_0^2 \hbar \omegac} {e \VG \lB^2 }\right].
\end{align*}

Then $T_0=2\EB/ \kB$ used in Ref.~\onlinecite{cohen2023universal} follows the general form
\begin{align*}
T_0 \propto \frac A {V_\mathrm{EW}-V_\mathrm{NS}} \,e^{-V_0/(V_\mathrm{EW}-V_\mathrm{NS})}
\end{align*}
where
$A= \kappa^2 E_C^2 (W_0/\lB)^2 / 2e\kB \approx 10^{3}\, \kappa^2 \,\si{K}\cdot\si{V}$
and
$V_0=4 \kappa^2 (W_0/\lB)^2 (\hbar \omegac)/e\approx 10^{2} \, \kappa^2 \,\si{V}$,
substituting $\VG = V_\mathrm{EW}-V_\mathrm{NS}$. 

This functional form generally matches the relationship between $T_0$ and $V_\mathrm{NS}$ found in Ref.~\onlinecite{cohen2023universal}, resembling an exponential in a neighborhood of $V_\mathrm{NS}$. The general scale can be made to fit with an appropriate choice of $\kappa$. Since this proportionality factor occurs squared in the exponential, it has a large effect on the scale of $T_0$.

\subsection{Connecting to spin-chain numerics}
We also want to relate $\EB$ to the spin-chain impurity $u$ used in the numerical simulations found in the main text. This impurity acts on a single site, giving a Fourier-transformed potential of
\begin{equation*} 
\tilde V(k)= \sqrt{ \frac 2 \pi } \frac{u \, \sin (ka_0/2)} {k}
\end{equation*}
where $a_0$ is the lattice constant. To calculate the sine-Gordon impurity $g$, we evaluate this at $\kF=\pi n/a_0$ where $n$ is the 1D per-site occupancy (measured to be approximately $0.7$ particles per site in our simulations). This gives a barrier energy of 
\begin{equation*} 
\EB = \frac{a u^2 \sin ^2(\pi n)}{2 \pi ^2 n^2 v}.
\end{equation*}
Using the known velocity of $v=0.5$, this expression shows a quadratic relationship between $\EB$ with a prefactor on the order of $10^{-1}$, assuming the refermionized cutoff $a$ to be on the order of 1 lattice site. 

We see this quadratic relationship in Fig.~\ref{fig:ac_conductance}(bottom) of the main text, which yields a prefactor of $0.55$. This value is roughly the same order as expected.

\section{Leads at different temperatures}
An interesting extension of our analysis occurs when the leads are at different temperatures $T_\pm$, which results
in a non-Fermi distribution for the $\psi(x)$ fermion given by the modified Fermi distribution $\tildenF (k)$.
This distribution is derived from noting that $\chiral\psi$ is an exponential of $\xi$ and 
therefore also of $\phi_\pm$ and observing that these bosonic fields have Gaussian distribution. 

The effective fermion field $\chiral\psi$ is expressed in terms of the chiral boson field $\chiral\xi(x)$, which relates to the original chiral fields as
\begin{align*}
    \chiral\xi(x) &= \frac \Gamma {\sqrt{K_-}} \phi_{-,\mathrm c}(x) + \frac \Gamma {\sqrt{K_+}} \phi_{+,\mathrm c}(x)
\end{align*}
where
\begin{align*}
    \phi_{\pm,\mathrm c}(x) &= \frac 1 2 \left[K_\pm^{-1/2} \phi_\pm(x) + K_\pm^{1/2} \int_0^x \mathrm dx'\, \Pi_\pm(x') \right].
\end{align*}
Using the canonical commutation relation $[\phi(x),\Pi(x')]=i \pi \delta(x-x')$ and the folded field definitions $\phi_\pm(x)=\phi(\pm x)$ and $\Pi_\pm(x)=\Pi(\pm x)$, we can show that the two chiral fields 
$\phi_{+,\mathrm c}$ and $\phi_{-,\mathrm c}$ commute with each other (though not with themselves at differing $x$). 
Therefore, products of effective fermions separate as
\begin{align*}
\left\langle \chiral\psi^\dagger(x_1)\, \chiral\psi(x_2) \right\rangle&= \left\langle e^{2i\chiral\xi(x_1)} \, e^{-2i\chiral\xi(x_2)} \right\rangle \\
&= \left\langle e^{(2\Gamma/\sqrt{K_+}) i\phi_{+,\mathrm c}(x_1)} e^{-(2\Gamma/\sqrt{K_+})i\phi_{+,\mathrm c}(x_2)}\right\rangle \left\langle e^{(2\Gamma/\sqrt{K_-})i\phi_{-,\mathrm c}(x_1)} e^{-(2\Gamma/\sqrt{K_-})i\phi_{-,\mathrm c}(x_2)} \right\rangle,
\end{align*}
assuming that far from the impurity $\phi_{-,\mathrm c}$ and $\phi_{+,\mathrm c}$ are independent.
Using the property of bosonic operators that $\langle \mathrm{exp} [\lambda \hat B]\rangle=\mathrm{exp} [\lambda^2 \langle {\hat B}^2 \rangle/2]$, we calculate
\begin{align}
    \left\langle \chiral\psi^\dagger(x_1)\, \chiral\psi(x_2) \right\rangle &= e^{-4\Gamma^2/K_+[\phi_{+,\mathrm c}(x),\phi_{+,\mathrm c}(x')]/2} e^{-4\Gamma^2/K_+\left\langle(\phi_{+,\mathrm c}(x_1)-\phi_{+,\mathrm c}(x_2))^2\right\rangle} \times \left(+\Longleftrightarrow -\right) \nonumber \\
&= \left\langle e^{2 i\phi_{+,\mathrm c}(x_1)}\; e^{-2i\phi_{+,\mathrm c}(x_2)}\right\rangle^{\Gamma^2/K_+} \left\langle e^{2i\phi_{-,\mathrm c}(x_1)} e^{-2i\phi_{-,\mathrm c}(x_2)} \right\rangle^{\Gamma^2/K_-} \nonumber \\
&= \left\langle \psi^\dagger_+ (x_1) \psi_+(x_2)\right\rangle^{\Gamma^2/K_+} \left\langle \psi^\dagger_- (x_1) \psi_-(x_2) \right\rangle^{\Gamma^2/K_-}.\label{eq:fermion-correlator}
\end{align}
where $\psi_\pm(x) = \mathrm{exp}[ 2i\phi_\pm(x) ]$. If $\psi_\pm$ is described by a Fermi-Dirac distribution $\nF$ with temperature $T_\pm$, these correlators are gives by the Fourier transform of $\nF(k)$:
\begin{equation*}
\left\langle \psi^\dagger_\pm (x+r) \psi_\pm(x)\right\rangle = -\frac{i\pi T_\pm}{\sinh \pi T_\pm x}
\end{equation*}
assuming translation invariance far from the boundary. Using this expression and Eq.~\ref{eq:fermion-correlator}, we find an effective distribution

\begin{align}
    \tildenF(k) &=\int \mathrm dx\, e^{i k x}\left(\frac{-i\pi T_+}{\sinh{\pi T_+ x}}\right)^{\Gamma^2/K_+}\left(\frac{-i\pi T_-}{\sinh{\pi T_- x}}\right)^{\Gamma^2/K_-}, \nonumber \\
    &=\int_0^\infty \mathrm dx \sin{(k x)} \, \Phi(x),
\end{align}
where 
\begin{equation*} 
 \Phi(x)=\left(\frac{T_\R}{\sinh{\pi T_\R x}}\right)^{\Gamma^2/K_\R}\left(\frac{T_\L}{\sinh{\pi T_\L x}}\right)^{\Gamma^2/K_\L}.
\end{equation*} 
This formula reduces to the Fermi distribution for $T_\L=T_\R$. 

The current, conductance and shot noise can now be computed by simply substituting $\tildenF$ for $\nF$. For the conductance, we can use the Eq.~\ref{eq:rAndreev} from the main text to get 
\begin{align*}
    G(V) &=\frac{e^2}{2 h} \left( 1-\frac \EB 2\int_0^\infty \mathrm dx \,x\,\Phi(x)\,e^{-\EB x}\cos{(V/2) x}\right).
\end{align*}
We plot the results of this function, numerically integrated with a Gauss-Kronrod quadrature formula, in Fig.~\ref{fig:G_V}. The result shows a small difference in the conductance when the leads are at different temperatures compared to the conductance with an effective single temperature.

\begin{figure}
\centering
\includegraphics[width=0.6\columnwidth]{G_V.pdf}
\caption{\label{fig:G_V} Voltage-dependent conductance for various temperatures demonstrating approximate equivalence between asymmetric and symmetric lead temperatures. Inset shows absolute difference between $T_\L=0$, $T_\R=2.4$ conductance and $T_\L=1.0$, $T_\R=1.0$.}
\end{figure}

\bibliography{bib.bib}